\documentclass[final,3p,times]{elsarticle}

\usepackage{graphicx}
\usepackage{amssymb}
\usepackage{amsthm}
\usepackage{amsmath}
\usepackage{tabularx}
\usepackage{multirow}
\usepackage{booktabs}
\usepackage{microtype} 
\usepackage[parfill]{parskip}
\usepackage{soul}
\usepackage{color}
\usepackage{cancel}
\usepackage{hyperref}
\usepackage{cleveref}
\AtBeginEnvironment{appendices}{\crefalias{section}{appendix}}
\usepackage[ruled,vlined]{algorithm2e}
\usepackage{listings}
\usepackage{caption}
\usepackage{subcaption}
\usepackage{array}
\usepackage{etoolbox}
\usepackage{appendix}

\begin{document}
\begin{frontmatter}
\title{Photo2CAD: Automated 3D solid reconstruction from 2D drawings using OpenCV}

\author[1]{Ajay B. Harish \corref{one}}
\ead{ajay.harish@manchester.ac.uk}
\author[1]{Abhishek R Prasad}
\cortext[one]{Corresponding author at: School of Engineering, University of Manchester, UK}
\address[1]{School of Engineering, University of Manchester, UK}

\begin{abstract}
This study showcases the utilisation of OpenCV for extracting features from photos of 2D engineering drawings. These features are then employed to reconstruct 3D CAD models in SCAD format and generate 3D point cloud data similar to LIDAR scans. Many historical mechanical, aerospace, and civil engineering designs exist only as drawings, lacking software-generated CAD or BIM models. While 2D to 3D conversion itself is not novel, the novelty of this work is in the usage of simple photos rather than scans or electronic documentation of 2D drawings. The method can also use scanned drawing data. While the approach is effective for simple shapes, it currently does not address hidden lines in CAD drawings. The Python Jupyter notebook codes developed for this purpose are accessible through GitHub.
\end{abstract}

\end{frontmatter}

\section{Introduction}
The growth of 3D modelling software revolutionised rapid 3D prototyping through Computer-Aided Design (CAD) while recent advances in machine learning (ML) has helped automate this process. Engineering designs employ 2D CAD drawings and 3D solid models for geometry representation, often using formats like \texttt{STEP, IGES} (open-source) and \texttt{SLDPRT} (proprietary) etc. These technologies and interpreters, facilitating format conversion with minimal data loss, has marked the move from the original paper drawings to computerised models. Today, numerous legacy 2D drawings remain from pre-software era, necessitating algorithms to transform them into solids models that can be directly used in manufacturing. This topic has been dealt as early as the 80's \cite{aldefeld1983a, preiss1984a, chen1988a, nagendra1988a, kargas1988a, gujar1989a, liu1994a}. The methods proposed to date can be broadly classified into (a) pattern recognition, (b) state transition, (c) decomposition, (d) graph-based (e) ML-driven. Among all, the most primitive and simplest approach is to use geometric entities, like points, edges and faces to convert them into equivalent 3D components. While such approaches are easier to implement, the extraction techniques cannot represent all forms of geometries.

A detailed literature review about the recent developments can be found in the work of \citet{shi2020a}. While the topic is not new, it remains of topical to core engineering domains where recent novelties in computer graphics can assist. An automated conversion of the 2D drawing to 3D CAD is not limited in interest to mechanical engineering and manufacturing but to other domains like biomedical etc. where image based data is the norm. This study leverages computer vision ideas to convert 2D engineering drawings into 3D CAD models. Using engineering views for reconstruction, it covers DXF, PDF, and camera-sourced drawings, adaptable for various objects through 3D CAD reconstruction from top, side, and front views. While currently limited to SCAD format due to its Boolean operation support, the method shows future promise for extension to use other formats, like STEP or IGES.

A Python Jupyter Notebook interface is developed to provide swift extraction and conversion of 2D CAD drawings (from PDF's, DXF files, or photos) into real-time 3D solid SCAD models. These SCAD models automatically transform into 3D point clouds, primed for further processing into 3D meshes for applications such as FEA analysis or manufacturing. The open-source Jupyter notebooks outlining these methods are accessible on GitHub. The subsequent sections presents theoretical and algorithmic explanations, followed by concise result summaries and conclusions.

While 2D drawing to 3D CAD conversion itself is not new, the novelty of this work remains that it leverages photos rather than scans or 2D electronic drawings for the conversion process. Thus,
\begin{itemize}
    \item The primary novelty of this work is the use of photo of a 2D drawing instead of scan for 3D conversion.
    \item A peripheral benefit also remains that if 2D drawings are lost but the product exists, then a photo of the three views can still be used to convert to 3D. However, this can be better achieved with other techniques including 3D scanners. This is not the primary novelty of this work but an additional benefit.
\end{itemize}

\section{Theoretical formulations}
\begin{figure}[!htb]
\centering
\includegraphics[width = \textwidth]{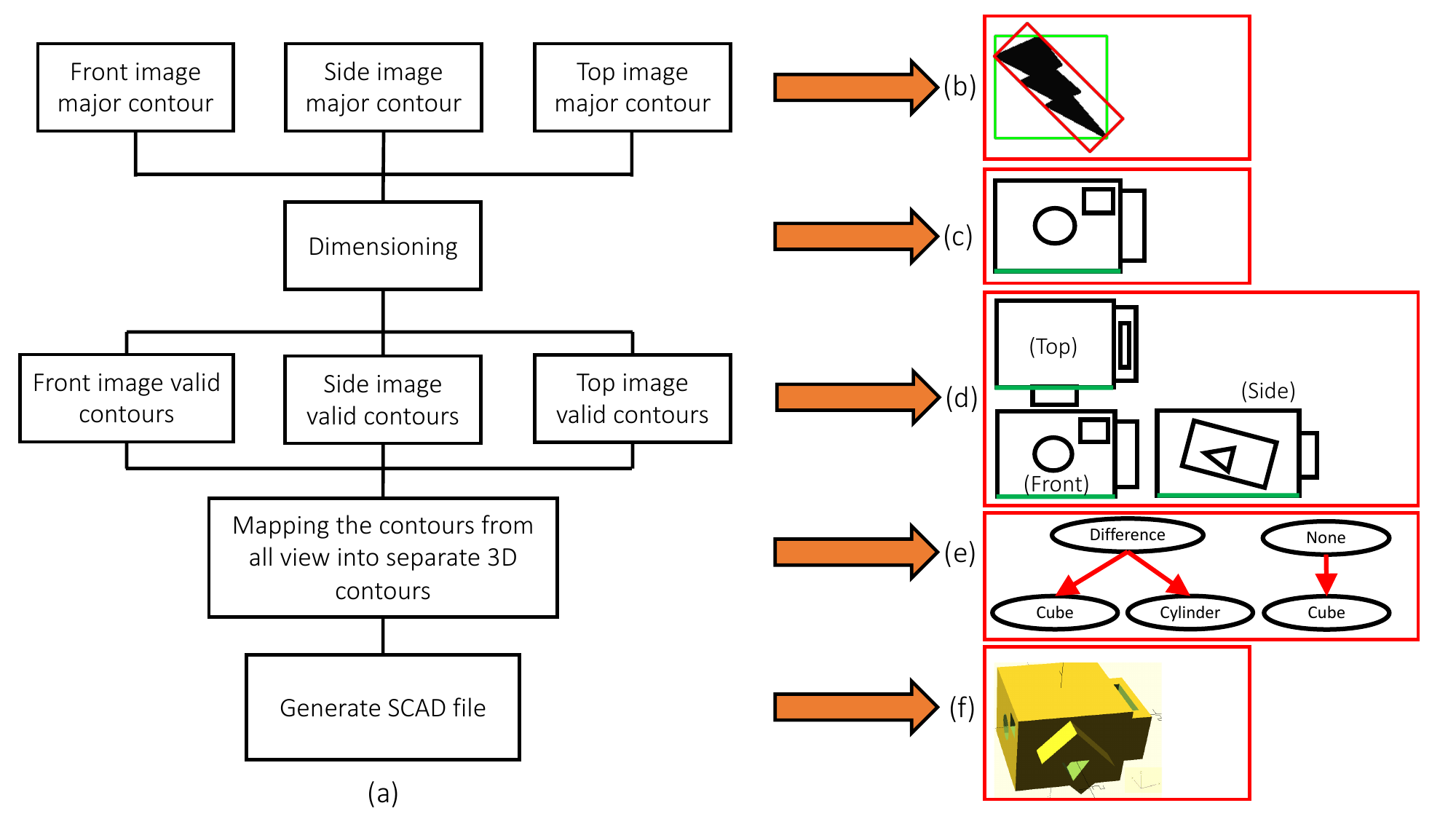}
\caption{(a) Flowchart describing the conversion of 2D view to 3D model; (b) Green rectangle shows the normal bounding rectangle while red rectangle shows the rotated bounding box that is an accurate fit (c) Dimensioning and extraction of length in one single view; (d) Extraction and dimensioning of different views; (e) Creation of binary tree for SCAD file; (f) Final SCAD geometry.}
\label{process flow}
\end{figure}

This work leverages OpenCV to perform feature identification, recognition, and conversion into 3D SCAD solids. It commences with the photograph of a 2D drawing or a scanned version. The overall flow is given in the \Cref{process flow}.
The reconstruction process consists of detecting outer boundaries, creating a bounding box, pointing locations relative to this box, and ultimately converting these points into a 3D CAD model in SCAD format.

\subsection{Contour detection}
The first step involves transformation of the image into a format conducive for feature discernment. While \textit{edges} and \textit{contours} are often used in common literature, they carry distinct implications within computer graphics community. Edges pertain to points where values undergo substantial variations relative to neighbouring points, encapsulating a localised concept. In contrast, contours denote enclosed curves derived from edges, delineating the boundaries of figures. A contour line signifies a curved demarcation portraying regions of identical values or intensities, effectively representing the holistic figure boundary. The OpenCV-computed contour manifests as a point array encompassing these contour lines, necessitating the sorting of contours by their areas to delineate the outer periphery of the 2D depiction. This process is initiated through a grayscale conversion of the image followed by contour rearrangement.

In order to extract image features, a bounding box is indispensable. Consequently, it becomes imperative to ascertain the center and extreme points of the contour. After converting the image to grayscale and isolating the contours, the OpenCV function \texttt{cv2.moments()} proves instrumental in obtaining a comprehensive dictionary of moment values computed for the contour. In computer vision, the moment signifies the centroid of an image, denoted as a weighted average of pixel intensities. This can be succinctly expressed as
\begin{equation}
M_{pq} = \int_{-\infty}^{\infty} \int_{-\infty}^{\infty} {x^p y^q f\left(x,y\right) \ dx dy}
\end{equation}
for $p,q = 0,1,2,\cdots$. This can be further adapted for the grayscaled image with pixel intensities $\mathbf{I}\left(x,y\right)$ to determine the raw image moments $M_{ij}$ as
\begin{equation}
M_{ij} = \sum_{x} \sum_{y} {x^i y^j \mathbf{I}\left(x,y\right)}
\end{equation}
The centroid of this image is then given to be
\begin{equation}
C_x = \frac{M_{10}}{M_{00}} \ \text{ and } \ C_y = \frac{M_{01}}{M_{00}}
\end{equation}

The next step in the determination of the ends of the figure includes the evaluation of the bounding box. As shown in \Cref{process flow}(b), the object of interest could be rotated. The area of the straight rectangle is not always the minimum. Hence, it would be necessary to consider both the extremum points and the object's rotation to evaluate the minimum area of the bounding rectangle. 

\subsection{Point location}
A signed distance function is used to determine a point's location in relation to the contour. This function returns +1 (for inside), -1 (for outside), 0 (on) depending on relative position of the point with respect to the contour.

\subsection{Conversion to 3D model}
The last step involved involves the recognition of 2D shapes for conversion into 3D solids. This work employs the Ramer–Douglas–Peucker algorithm \cite{ramer1972a,douglas1973a} to convert the curve into a set of points. In this work six shapes: triangle, square, rectangle, pentagon, hexagon, and circle are considered. A complete listing of the code is available in Section 1.5 of supplementary materials.

Finally, the evaluated shapes are converted into SCAD format using the basic shapes: sphere (\texttt{r =radius}); cube (\texttt{size = [x,y,z], center = true/false}), where the center lies at $(0,0)$; cylinder (\texttt{h,r1,r2,center, fn : fixed number of fragments}). This is further followed by Boolean operations, namely \texttt{union()} and \texttt{difference()} leading to complex shapes and objects. A sample of such union is shown in \Cref{process flow}(f).

\subsection{Algorithmic implementation}
The algorithmic implementation will discuss the methods involved in dimensioning, identification of shapes, translation and rotations of objects, combining multiple parts. The above steps are individually performed for all the views to finally create a combined tree for the SCAD file. This is followed by a 3D point cloud generation. The algorithmic implementation is broken into five stages and the algorithms themselves are provided in the supplementary materials.

\subsubsection{Convention}
The following convention is considered for the image processing:
\begin{itemize}
    \item The front view is pointed along the $y$-direction. The translation in $z$ and $x$-axis and rotation about the $y$-axis are obtained by image processing.
    \item The side view is towards -x direction. Since one is viewing from the $y$-axis for image processing, all the shapes recognised from the side-view are rotated by $90^{\circ}$ about the $z$-axis. Viewing the shapes from the $x$-axis, the rotations about $x$-axis and translation along $y$- and $z$- axis are also obtained.
    \item The top view is oriented towards the $z$-axis. Here the rotations about $z$-axis and translations about $x$- and $y$-axis are extracted.
\end{itemize}

\subsubsection{Dimensioning of objects}
The user-provided image is used in the determination of the pixel densities and object length. If the images containing the different views have different pixel densities, it would difficult to determine the actual dimensions of the shape accurately. Thus, as shown in \Cref{process flow}(c), the user also defines the dimension of the highlighted line of the object, which is used in the determination of the object's length to pixel ratio. The overall process of determination of the outermost contour is as shown in algorithm 1 in supplementary materials. 

\subsubsection{Linked parent-child lists}
The shape details are further decoupled and stored in a linked list as
\begin{itemize}
\label{parts of shape}
    \item node object - Class object containing details of a shape for constructing 3D model for that shape
    \item tuple (\texttt{[cX,cY]}) - center of contour
    \item contour from image processing / index of the parent-child
    \item height of the horizontal rectangle enclosing the contour($h$)
    \item width of the horizontal rectangle enclosing the contour($w$)
\end{itemize}
As shown in \Cref{process flow}(c), the views likely have one object inside another, here a circle and rectangle inside a bigger rectangle. Thus, a parent-child relation grouping, through the shape list, is a necessity. Each shape list in each view can be organised as follows
\begin{equation}
\label{object parent-child}
\texttt{objects}= [\{ \texttt{parent}_{1}, \texttt{child}_{11}, \texttt{child}_{12}..\},\{\texttt{parent}_{2}, \texttt{child}_{21}, \texttt{child}_{22}..\},....] 
\end{equation}
where the \texttt{objects} is a list of lists, with each list outlining the parent followed with child details. Each parent and child have all the details of the shape list that are only visible from that particular view. For example, in \Cref{process flow}(c), the rectangle is a parent of the circle and thus, the list of objects in the front view will contain
\begin{equation}
\texttt{object}_{front} = [\texttt{rectangle}_{detail-list}, \texttt{Circle}_{detail-list} \cdots]
\end{equation}
Since these parent-child lists are limited to a particular view, they do not provide a complete perspective of the 3D shape. For example, the front view consists of information about the height and the length but does not provide any information about the object's breadth. This is available in the side view. Thus, once all the views have been parsed, a final parent-child list containing all details of the 3D shape is created. 

\subsubsection{Parent-child list for each view}
Algorithm 2, in supplementary materials, describes the process of creating the parent-child lists for a particular view. Firstly, all the contours recognised are arranged in the descending order of the area occupied by the contour. Further on, looping through each contour, one can determine the shape of the contour. As in the above example, this would translate to the determination of the rectangle's length and height, the radius of circles, etc. Additionally, the center of the largest contour is also considered to calculate the child object's translation and rotations in relation to the parent object. Based on the recognised shape, a node class object is defined as a cube for square and rectangle, and cylinder for triangle, pentagon, hexagon, and circle.

\subsubsection{Re-arrangement of parent-child lists in each view}
This entire process is outlined in algorithm 3 in the supplementary materials. Prior to the SCAD file's consolidation and generation, it is necessary to integrate the parent-child lists from each view. The front view list is stored in a descending order of the $x-$ coordinate of the parent's center obtained from image processing. The other two views are stored in ascending order. For example, if the center coordinates of the $\texttt{parent}_{2}$ are $(6,5)$ and that of $\texttt{parent}_{1}$ are $(3,8)$, then the object lists created as shown in \Cref{object-list02}.
\begin{equation}
  \texttt{object}_{front}=[[\texttt{parent}_{2}, \texttt{child}_{21}, \texttt{child}_{22}..],[\texttt{parent}_{1}, \texttt{child}_{11}, \texttt{child}_{12}..]]  
  \label{object-list02}
\end{equation}

\subsubsection{2D views to SCAD models}
Once the list of parent-child objects has been assimilated from all the views, it is essential to separate the duplicates before combining them into a single shape. Here, the front view is considered a primary list, and the shapes from other views are added and updated using the algorithm 4 in supplementary materials described below.

In order to assimilate the views and convert them into a 3D CAD model, different cases are considered. For each parent-child object list available in the front view, their counterpart with the same shape is searched in the side view. This is achieved by comparing the height of the shape stored in the shape detail list. In order to allow for variations, a small error, called \texttt{roundOffApprox} is allowed. If the shape of interest is present in the side view but is not a parent, then the shape is being extruded. Thus, a union operation is associated with this shape. Alternatively, if a counterpart exists in the side view, then the rest of the shape in the list are added as holes on the parent. This is always the case since the parent-child lists are already re-arranged in the descending order of the center $x$- coordinate of the parents. Thus, this automatically allows one to consider all the shapes that require an extrusion at the first pass. If the front view has a cylinder, then the missing dimensions are updated from the side view. Alternatively, if the front view has the form of a cube, then two possibilities need to be considered in the side view, namely:
\begin{itemize}
\item \textbf{Cube:} The search continues in the top view where there are two possibilities that the shape is a cylinder or a cube. In the former case, the front shape is replaced with a cylinder with proper translation and all the missing dimensions and index in the parent-child list is stored as a part of the top view shape details. In the latter case, the front shape is replaced by a cube along with any rotation about the $z-$ and/or $x-$ axis. In all the cases, relevant missing dimensions are updated.
\item \textbf{Cylinder:} If this is a cylinder, then the front view cube is replaced by side view cylinder with proper height and translation
\end{itemize}
    
However, for the shapes not visible in the front view but visible in the side view similar iteration over the parent-child list of the side view is considered. If the shape is not already considered in the earlier iterations, the search for its counterpart in the top view is initiated. Similar iteration between the side and top views are considered to determine all the objects outlined in the side view. 
    
Finally, if the shape is only visible in the top view, this is likely a hole visible only in the top view. This also means that the hole's parent should be at least visible in the front or side view. Else, that shape can be ignored. If the parent was visible, then the parent-child index will tell us where to add these holes in the parent-child list.

\begin{figure}[!htb] 
    \centering
    \includegraphics[width=\linewidth]{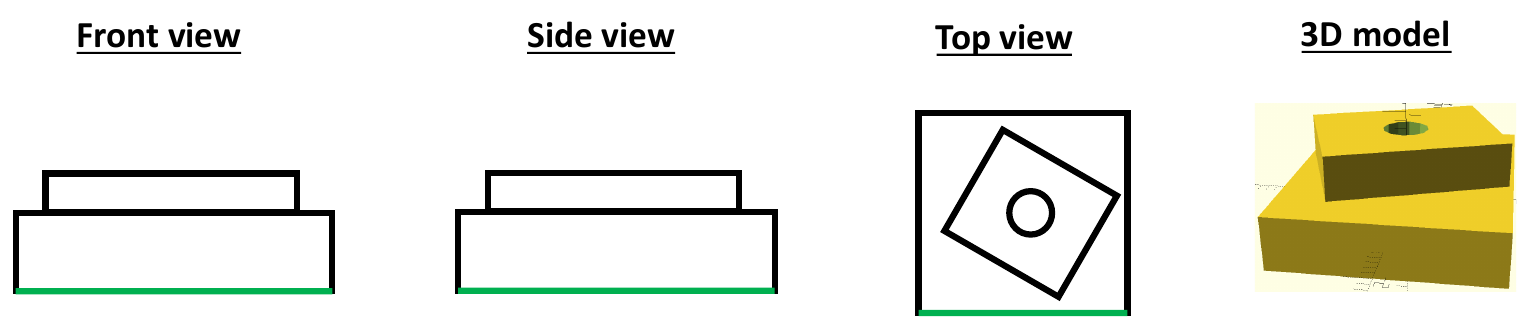} 
  \caption{Simple example demonstrating the different views and the final SCAD model generated from these views}
  \label{ViewsSCAD}
\end{figure}

Once the final parent-child lists have been set up, they are compressed into a single list. The Boolean operation for the parent objects will be \texttt{Null} while that of the children objects are either \texttt{Union} or \texttt{Difference}. This final list's transformation to a 3D SCAD file is outlined in the algorithm 5 in supplementary materials. A binary tree is created for each parent-child. Here, each shape node class object is a node in the binary tree, and each shape node is a child node of an empty node class object. The empty node class only holds the string of SCAD syntax, including the operation between the two children. A sample tree created is as shown in \Cref{process flow}(e). Here, the example considered is a simple shape shown in \Cref{ViewsSCAD}.

\section{Results and discussions}
The developed model can recognise several 2D shapes, including triangle, square, rectangle, pentagon, and hexagon. However, 3D models with objects rotated about multiple axes are not supported. Additionally, the translations considered are relative to the outermost contour in each view. Thus, the outermost contour in each view needs to represent the same 2D shape. In order to demonstrate the applicability of the developed method, simple geometries of interest are considered and shown in \Cref{example}. In \Cref{example}, the primary dimensions of the outermost contour, extracted from each view, is marked and shown in green. The table also shows the final SCAD model created. The shapes demonstrated here are simple but the developed methodology using OpenCV is generic and can be extended for more complicated shapes and models.

\subsection{Converting the 3D SCAD models to point cloud}
\begin{figure}[!htb] 
    \centering
    \includegraphics[width=\linewidth]{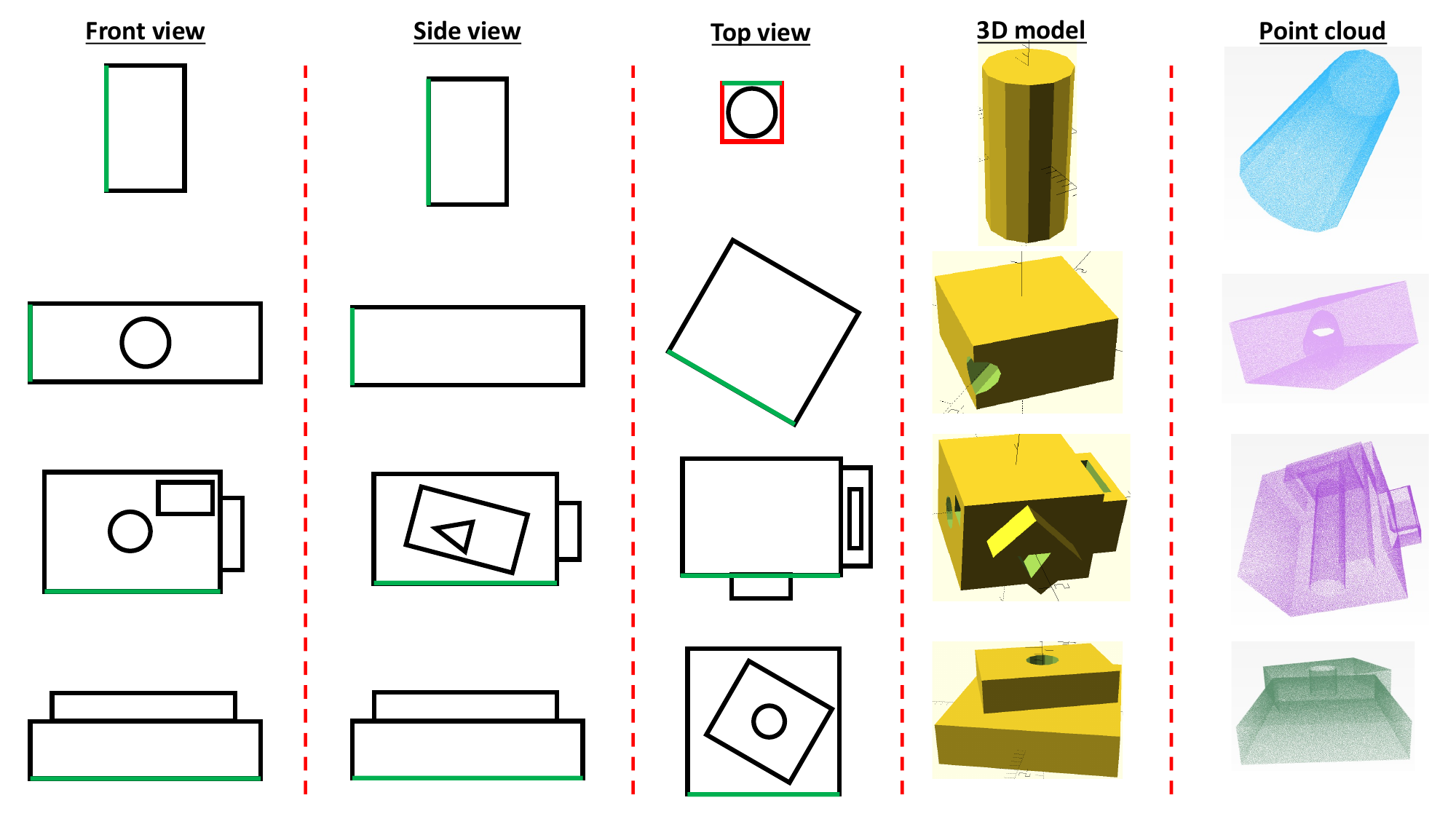} 
  \caption{Examples demonstrating the identification, dimensioning and conversion of the different views into 3D model and subsequent conversion into point cloud.}
  \label{example}
\end{figure}

While SCAD models are useful, they do not find industry applications yet. SCAD framework provides tools required to convert the models to more industry suitable formats like \texttt{STEP and IGES} etc. Further on, it is more useful to convert these SCAD models to 3D point clouds that can be used for automatically generating high-quality meshes for numerical simulations like finite element analysis and computational fluid dynamics. In this regard, the generated SCAD file is first converted to \texttt{STL} format. The open3d method proposed by \citet{yuksel2015a} and implemented in the \texttt{Open3d} library is directly incorporated to determine the points and generate a 3D point cloud. \Cref{example} also shows the point clouds generated for the objects considered.
\section{Conclusions and Future Work}
The automatic conversion of 2D engineering drawings to 3D CAD models is a well-established topic, its relevance endures. This study employs cutting-edge computer vision advancements to create an algorithm for seamless automated conversion, specifically generating SCAD geometries from photos of 2D counterparts. Although the considered geometries remain simple, the methodology's potential can be amplified by the prospect of integrating shape recognition via neural networks. This enhancement can seamlessly integrate into the provided python Jupyter notebook. While not a comprehensive solution, this work presents a novel avenue for synergising contemporary computer graphics and vision advancements with core engineering applications.

\section*{Acknowledgements}
This work has received support from multiple sources. The authors would like to thank and acknowledge the support received from (a) The School of Engineering at the University of Manchester through startup funds; (b) Engineering and Physical Sciences Research Council (EPSRC) Impact Acceleration Account via The University of Manchester [IAA 391]; (c) Isaac Newton Institute for Mathematical Sciences for support and hospitality during the programme Uncertainty Quantification and Stochastic Modelling of Materials when work on this paper was undertaken. This work was supported by EPSRC Grant Number EP/R014604/1.

\bibliographystyle{elsarticle-harv}
\bibliography{Reference}

\section*{CRediT authorship contribution statement}
\textbf{Abhishek R Prasad:} Software, Methodology, Validation, Formal analysis, Investigation, Data Curation, Visualisation; \textbf{Ajay B Harish:} Conceptualisation, Methodology, Resources, Writing - Original draft, Project administration, Funding acquisition.

\section*{Data availability}
The codes from this paper are available through GitHub at \url{https://github.com/Ajay-Research-Lab/Photo2CAD}.

\end{document}